\documentstyle[12pt,epsf]{article}
\renewcommand{\baselinestretch}{1.0}

\newcommand{\be}{\begin{equation}}
\newcommand{\ee}{\end{equation}}
\begin{document}
\topmargin 0pt
\oddsidemargin=-0.4truecm
\evensidemargin=-0.4truecm
\renewcommand{\thefootnote}{\fnsymbol{footnote}}

\newpage
\setcounter{page}{1}
\begin{titlepage}     
\vspace*{-2.0cm}
\begin{flushright}
\vspace*{-0.2cm}
\end{flushright}
\vspace*{0.5cm}

\begin{center}
{\Large \bf Solar Neutrinos: Global Analysis  with 
Day and Night Spectra from SNO}
\vspace{0.5cm}

{P. C. de Holanda$^{1}$ and  A. Yu. Smirnov$^{1,2}$\\
\vspace*{0.2cm}
{\em (1) The Abdus Salam International Centre for Theoretical Physics,  
I-34100 Trieste, Italy }\\
{\em (2) Institute for Nuclear Research of Russian Academy 
of Sciences, Moscow 117312, Russia}

}
\end{center}

\vglue 0.8truecm
\begin{abstract}

We perform global analysis of the solar neutrino data including the
day and night spectra of events at SNO.  In the context of two active
neutrino mixing, the best fit of the data is provided by the LMA MSW
solution with $\Delta m^2 = 6.15 \cdot 10^{-5}$ eV$^2$, $\tan^2 \theta
= 0.41$, $f_B = 1.05$, where $f_B$ is the boron neutrino flux in units
of the corresponding flux in the Standard Solar Model (SSM).  At
$3\sigma$ level we find the following upper bounds: $\tan^2 \theta <
0.84$ and $\Delta m^2 < 3.6 \cdot 10^{-4}$ eV$^2$.  From
$1\sigma$-interval we expect the day-night asymmetries of the charged
current and electron scattering events to be: $A_{DN}^{CC} =
3.9^{+3.6}_{-2.9}\%$ and $A_{DN}^{ES} = 2.1^{+2.1}_{-1.4}\%$.  The only
other solution which appears at $3\sigma$-level is the VAC solution
with $\Delta m^2 = 4.5 \cdot 10^{-10}$ eV$^2$, $\tan^2 \theta = 2.1$
and $f_B=0.75$.  
The best fit point in the LOW region, with $\Delta
m^2 = 0.93 \cdot 10^{-7}$ eV$^2$ and $\tan^2 \theta = 0.64$, is
accepted at $99.95\%$ ($3.5\sigma$) C.L. .  The least $\chi^2$ point
from the SMA solution region, with $\Delta m^2 = 4.6 \cdot 10^{-6}$
eV$^2$ and $\tan^2 \theta = 5 \cdot 10^{-4}$, could be accepted at
$5.5\sigma$-level only.  In the three neutrino context the influence
of $\theta_{13}$ is studied. We find that with increase of
$\theta_{13}$ the LMA best fit point shifts to larger $\Delta m^2$,
mixing angle is practically unchanged, and the quality of the fit
becomes worse. The fits of LOW and SMA slightly improve. Predictions
for KamLAND experiment (total rates, spectrum distortion) have been
calculated.

\end{abstract}
\centerline{Pacs numbers: 14.60.Lm 14.60.Pq 95.85.Ry 26.65.+t} 
\end{titlepage}
\renewcommand{\thefootnote}{\arabic{footnote}}
\setcounter{footnote}{0}
\renewcommand{\baselinestretch}{0.9}

\section{Introduction}

The SNO data \cite{sno,sno1,sno-nc,sno-dn,sno-how}  is the breakthrough in
a long story of the 
solar neutrino problem. With high confidence level we can claim 
that solar neutrinos undergo the flavor conversion
\be  
\nu_e \rightarrow \nu_{\mu}, \nu_{\tau}~~{\rm or/and} ~~ \bar\nu_{\mu},
\bar\nu_{\tau}.  
\ee
Moreover, non-electron neutrinos compose larger part of the 
solar neutrino flux at high energies (also partial conversion to sterile 
neutrinos is not excluded). 
The main issue now is to identify the {\it mechanism} of neutrino
conversion. 

There are several important pieces of new information 
from the recent SNO publication \cite{sno-nc,sno-dn,sno-how}: 

1. Measurements of the energy spectra with low threshold 
(as well as angular distribution) of
events allow to extract information on the 
neutrino neutral current (NC),  charged current (CC), as well as  
electron scattering (ES) event rates. In particular, in assumption of 
absence of distortion, one gets for the ratio of the
NC/CC event rates: 
\be
\frac{NC}{CC} = 2.9 \pm 0.4 
\label{nc/cc}
\ee    
which deviates from 1 by  about $5\sigma$. 
  
2. Measurements of the day and night energy spectra 
allow to find the D-N asymmetries of different  classes 
of events.  Under constraint that total flux has no D-N  asymmetry 
one gets for CC event rate~\cite{sno-dn}  
\be
A^{CC}_{DN} = 7.0 \pm 4.9 ~^{+1.5}_{-1.4}  \%  .        
\label{asym}
\ee
  
3. No substantial distortion of the neutrino energy spectrum
has been found. 

4. Solutions of the solar neutrino problem based on pure 
active - sterile conversion, $\nu_e \rightarrow \nu_s$,
are strongly disfavored. 

These results further confirm earlier indications 
of $\nu_{\mu}-, \nu_{\tau}-$ appearance 
from comparison of fluxes determined from the charged current 
event rate in the SNO detector \cite{sno,sno1}, and 
the $\nu e-$scattering event rate obtained by the 
Super-Kamiokande (SK) collaboration~\cite{SK,smy1,smy2}. 

Implications of the new SNO results for different solutions 
(see \cite{bks-10-cor,bks01} for earlier studies)   
can be obtained immediately by
comparison of the results (\ref{nc/cc},  \ref{asym})  with predictions
from the best fit points of different solutions
\cite{barger,fogli,valencia,calcutta,plamen,gago,torr}. 
In particular,  for the  LMA solution the best fit prediction   
$NC/CC = 3.3$ (for lower threshold)~\cite{plamen} is 
slightly higher than (\ref{nc/cc}). So, new results
should move the best fit point to larger values of mixing angles. 
The expected day-night asymmetry in the best fit point, $\sim 6\%$,    is
well within the interval (\ref{asym}). Clearly new data further favor this
solution.

For LOW solution:
$NC/CC = 2.4$~\cite{plamen}  in the best fit point, 
which is $1 \sigma$ (experimental) lower than  the
central SNO value. 
The expected asymmetry was lower than (\ref{asym}), 
therefore this solution is somewhat less favored, and
SNO tends to shift the allowed region to smaller values of $\theta$ which
correspond to smaller survival probability.

Implications of new SNO results have been studied 
in~\cite{barger-a,bahcall-a,dproy,strumia,milano}. 
In this paper we continue this study.
We perform global analysis of all available data
including the SNO day and night energy spectra of 
events, and the latest data from Super-Kamiokande and SAGE. 
We identify the most plausible solutions and study their properties.

The paper is organized as follows: In Section~2 we describe features of 
our analysis. In Section~3 we present results of the 
$\chi^2-$ test, 
and construct the pull-off diagrams for various observables. 
In Section~4 we determine the regions 
of  solutions and describe their properties. In Section~5 we consider  the
effect of $\theta_{13}$ on the solutions. In Section~6 we study the
predictions to KamLAND for the parameters given by the found solutions. 
The conclusion is given in Section~7.

\section{Global analysis of the solar neutrino data}

In this section we describe main ingredients of our analysis. 
We follow the procedure of analysis developed in  previous
publications \cite{bks-10-cor,bks01,plamen,gago,fogli95}.

\subsection{Input data}

We use the following set of the experimental results: 

1) Three rates (3 d.o.f.): 

(i) the $Ar-$production rate $Q_{Ar}$ measured by the Homestake 
experiment \cite{Cl}, 

(ii) the $Ge-$production rate, $Q_{Ge}$, from 
SAGE \cite{sage}, 

(iii) the combined $Ge-$production rate from GALLEX
and GNO \cite{gno}.

2) The zenith-spectra 
measured by Super-Kamiokande \cite{SK} during 1496 days of 
operation. The data consists of  
8 energy bins with 7 zenith angle bins in each, except for the first and last
energy bins, which makes 44 data points.
We use the experimental errors given in~\cite{smy1} and we treat the
correlation of systematic uncertainties as in~\cite{gago}.
Following the procedure outlined in \cite{bks01} we do not include the
total rate of events in the SK detector, which is not independent from
the spectral data.

3) From SNO,  we use the day and the night energy spectra 
of all events \cite{sno-how}.  We follow procedure described in 
\cite{sno-how}. Additional information on how to treat the systematic
uncertainties was given by~\cite{private}. 

Altogether there are 81  data points.

\subsection{Neutrino Fluxes}

All solar  neutrino fluxes (but the boron neutrino flux) are taken
according to SSM BP2000 \cite{ssm}. 
We use the boron neutrino flux as free parameter. 
We define dimensionless parameter  
\be
f_B \equiv \frac{F_B}{F_B^{SSM}}, 
\ee
where the SSM  boron neutrino flux is taken to be 
$F^{SSM}_B = 5.05 \cdot 10^{6}$ cm$^{-2}$ s$^{-1}$.  
For the $hep-$neutrino flux we take fixed value  
$F_{hep} = 9.3 \times 10^{3}$ cm$^{-2}$ s$^{-1}$ \cite{ssm,hepfl} .

\subsection{Neutrino mixing and conversion}

We perform analysis of data in terms of 
mixing of two flavor neutrinos and three flavor neutrinos. 
 
In the case of two neutrinos  there are two  oscillation parameters:  
the mass squared difference,  $\Delta m^2$, and the
mixing parameter $\tan^2\theta$. 
So, we have three fit parameters: $\Delta m^2$, $\tan^2\theta$, $f_B$, 
and therefore 81(data points) - 3 = 78 d.o.f.

In the case of three neutrino mixing we adopt  the mass scheme 
which explains the solar and the atmospheric neutrino data. 
In this scheme the mass eigenstates $\nu_1$ and $\nu_2$ are splitted by
the solar $\Delta m^2_{12}$, whereas 
the third mass eigenstate, $\nu_3$, is separated by larger mass split 
related to the atmospheric $\Delta m^2_{13}$. 
Matter effect influences very weakly  mixing (flavor content)  
of  the third mass eigenstate. 
The effect of third neutrino is reduced then to the averaged vacuum
oscillations. In this case, the survival probability equals 
\be
P_{ee} = \cos^4 \theta_{13} P_{ee}^{(2)} +  \sin^4 \theta_{13},  
\label{surv3}
\ee
where $\sin \theta_{13} \equiv U_{e3}$ describes the mixing of electron
neutrino in the third mass eigenstate and  
$P_{ee}^{(2)}$ is the two neutrino oscillation probability  characterized
by  $\tan^2 \theta_{12}$, $\Delta m_{12}^2$ and the effective
matter potential reduced by factor $\cos^2 \theta_{13}$ 
(see e.g. \cite{three,three1} for previous studies).   

In general, in the three neutrino case the fit parameters are 
$\tan^2 \theta_{12}$, $\Delta m_{12}^2$,  $\sin \theta_{13}$ and 
$f_B$. However, here for illustrative purpose we take 
fixed value of $\theta_{13}$ near its upper bound.  
So, the number of degrees of freedom  
is the same as in the two neutrino case.

\subsection{Statistical analysis}

We perform the $\chi^2$ test of various oscillation solutions by
calculating
\be
\chi^2_{global} = \chi^2_{rate} +   \chi^2_{SK} + \chi^2_{SNO}, 
\label{chi-def}
\ee
where $\chi^2_{rate}$, $\chi^2_{SK}$  and  $\chi^2_{SNO}$
are the contributions
from the total rates, from the Super-Kamiokande zenith  spectra
and the SNO day and night spectra correspondingly. Each of 
the entries in Eq.(\ref{chi-def}) is the function of three
parameters ($\Delta m^2$, $\tan^2\theta$, $f_B$). 

Some details of treatment of the systematic errors are given in the Appendix. 

The uncertainties of contributions from different components of the solar neutrino flux (pp-, Be-, B- etc.) to
$Ge$-production rate due to uncertainties  of the cross-section of the detection reaction $\nu_e -  Ga$ correlate. 
Similarly,   uncertainties of contributions  to 
$Ar$ production rate due to uncertainty in $\nu_e -  Cl$ cross-section correlate. 
Following \cite{sterile} we have taken into account these correlations.

\subsection{Cross-checks. Comparison with other analysis}

We have checked our results performing two additional fits: 

1) To  check our treatment of the SK data we have  performed  
global analysis taking from SNO only the CC-rate. 
That corresponds to the analysis done in~\cite{smy1,smy2}. 
We get very good agreement of the results.

2) To check our treatment of the latest SNO data we have performed  
analysis using the  day and night spectra from SNO, as in 
\cite{sno-dn}. We have reproduced the results  of paper \cite{sno-dn}  
with a good accuracy.

Our input set of the data differs from that used in other analyses: We
include more complete and up-dated information.  SNO \cite{sno-dn} 
uses the SK day and night spectra measured
after 1258 days. In contrast, we use preliminary SK zenith spectra
measured during 1496 days. In~\cite{bahcall-a} the NC/CC ratio and the 
D-N asymmetry at SNO where included in the analysis. The analysis done
by Barger et al~\cite{barger-a} uses the same data set we do.

\section{$\chi^2$ test}

In this section we describe the results of fit for two neutrino
mixing. 

In Table \ref{Table1}  we show the best fit values of parameters 
$\Delta m^2$,  $\tan^2\theta$, $f_B$ for different solutions 
of the solar neutrino problem. We also give the corresponding 
values of $\chi^2_{min}$ and the goodness of the fit.    

The absolute $\chi^2$ minimum, $\chi^2 = 65.2$ for 78 d.o.f., 
is in the LMA region. The vacuum oscillation is the next best. It,
however, requires $\sim30\%$ lower boron neutrino flux. The LOW
solution has slightly higher $\chi^2$. The SMA gives a very bad fit.

\begin{table}
\caption[]{Best-fit values of the parameters $\Delta m^2$, 
$\tan^2\theta$ and  $f_B$, as well as 
the minimum $\chi^2$ and
the corresponding g.o.f. for various global solutions. 
The number of degrees of freedom is  78.} 
\vskip 0.5cm
\begin{tabular}{l c c c c c}
\hline
Solution & $\Delta m^2$/$\rm eV^2$ &  $\tan^2\theta$ & $f_B$ &
$\chi^2_{min}$ & g.o.f. \\ 

\hline

LMA      & $6.15\times 10^{-5} $ &     0.41   &  1.05 &   65.2  & 85\%\\

VAC      & $4.5\times 10^{-10} $ &  2.1   & 0.749 & 74.9 & 58\% \\

LOW      & $0.93\times 10^{-7} $ &   0.64 &  0.908   & 77.6 & 49\%\\
 
SMA      & $4.6\times 10^{-6} $ &  $0.5 \times 10^{-3}$ & 0.57 &  99.7 &
4.9\%\\

\hline 

\end{tabular}
\label{Table1}
\end{table}   

\begin{figure}[ht]
\centering\leavevmode
\epsfxsize=1.\hsize
\vspace{5cm}
\epsfbox{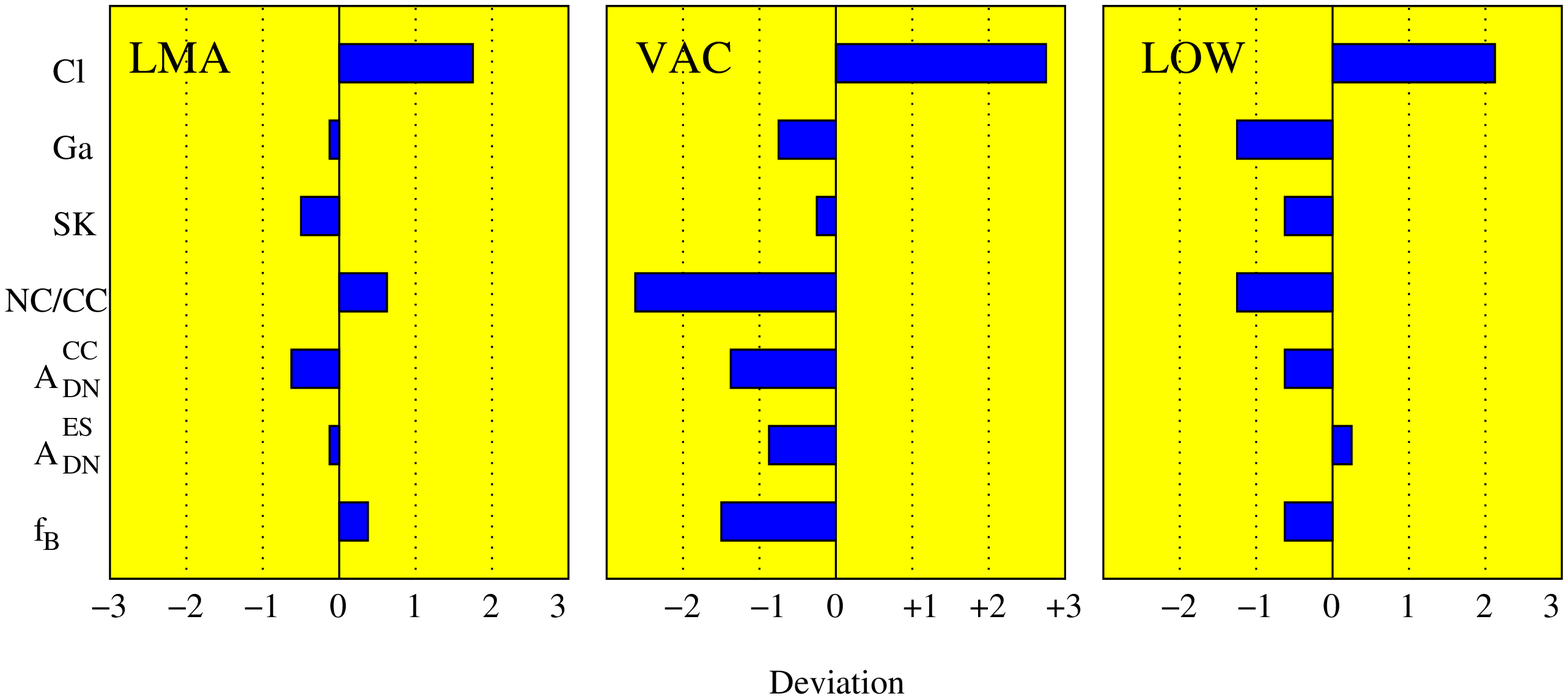}
\caption{
Pull-off diagrams for global solutions. Shown are deviations 
of predictions from experimentally measured values 
for the $Ar-$production rate,  
$Ge-$production rate, SK rate, the day-night asymmetries at SK 
and SNO.  
The pull-offs  are expressed  in the units of 1 standard deviation,  
$1 \sigma$.}
\label{pulloff} 
\end{figure}

In order to check the quality of the fits we have calculated
predictions for the available observables in the best fit points of
the global solutions (see Table 1).
Using these predictions we have constructed the ``pull-off" diagrams
(fig.~\ref{pulloff}) which show deviations, $D_K$, of the predicted
values of observables $K$ from the central experimental values
expressed in the $1\sigma$ unit:
\be
D_K \equiv \frac{K_{bf} - K_{exp}}{\sigma_K}, ~~~~ 
K \equiv Q_{Ar},~Q_{Ge},~ NC/CC,~ R_{\nu e},~ A_{DN}^{SK}, A_{DN}^{CC}.   
\label{pull}
\ee 
Here $\sigma_K$ is the one sigma standard deviation for a given
observable $K$. $R_{\nu e}$ is the reduced total rate of events at
SK. We take the experimental errors only: $\sigma_K =
\sigma_K^{exp}$.

According to Fig.~\ref{pulloff} only the LMA  solution does not 
have strong deviations  of predictions from the experimental results. 
LOW and VAC solutions give  worse fit to the data.\\

\section{Parameters of  solutions}

We define the solution regions by 
constructing the contours of constant (68, 90, 95, 99, 99.73~\%)
confidence level with respect of the
absolute minimum in the LMA region.  
Following the same procedure
as in \cite{bks01},  for each point in the $\Delta m^2$, $\tan^2 \theta$
plane we find 
minimal value $\chi^2_{min}(\Delta m^2, \tan^2 \theta)$
varying $f_B$.
We define the contours of constant confidence level by the condition
\be
\chi^2_{min}(\Delta m^2, \tan^2 \theta) = \chi^2_{min} (LMA) +
\Delta \chi^2~,
 \label{delta}
\ee
where $\chi^2_{min} (LMA) = 65.2$ is the absolute minimum in the LMA
region and $\Delta \chi^2$ is taken for two degrees of freedom.

\subsection{LMA}

Recent SNO data further favors the LMA MSW solution (see e.g. \cite{lma99}).  
In the best fit point we get
\be
\Delta m^2 =  6.15 \cdot 10^{-5}~ {\rm eV}^2, 
~~~\tan^2 \theta = 0.41, ~~~f_B = 1.05 . 
\label{bfparam}
\ee 
The value of $\Delta m^2$ is slightly higher than 
that found in the SNO analysis and higher than in our previous
analysis~\cite{plamen}. 
The shift is mainly due to updated SK results which show smaller 
D-N asymmetry than before. Large SNO asymmetry which would push 
$\Delta m^2$  to smaller values is still statistically insignificant.  
The mixing angle is shifted to larger values (in comparison with previous 
analysis) due to smaller ratio of the NC/CC event rates. 
The boron
neutrino flux is 5\% higher than central value in the SSM: 
$F_B = f_B \cdot F_B^{SSM} = 5.32 \cdot 10^6$ cm$^{-2}$ s$^{-1}$ 
being however within $1\sigma$ deviation and well in agreement with 
SNO measurements.

\begin{figure}[ht]
\centering\leavevmode
\epsfxsize=.8\hsize
\epsfbox{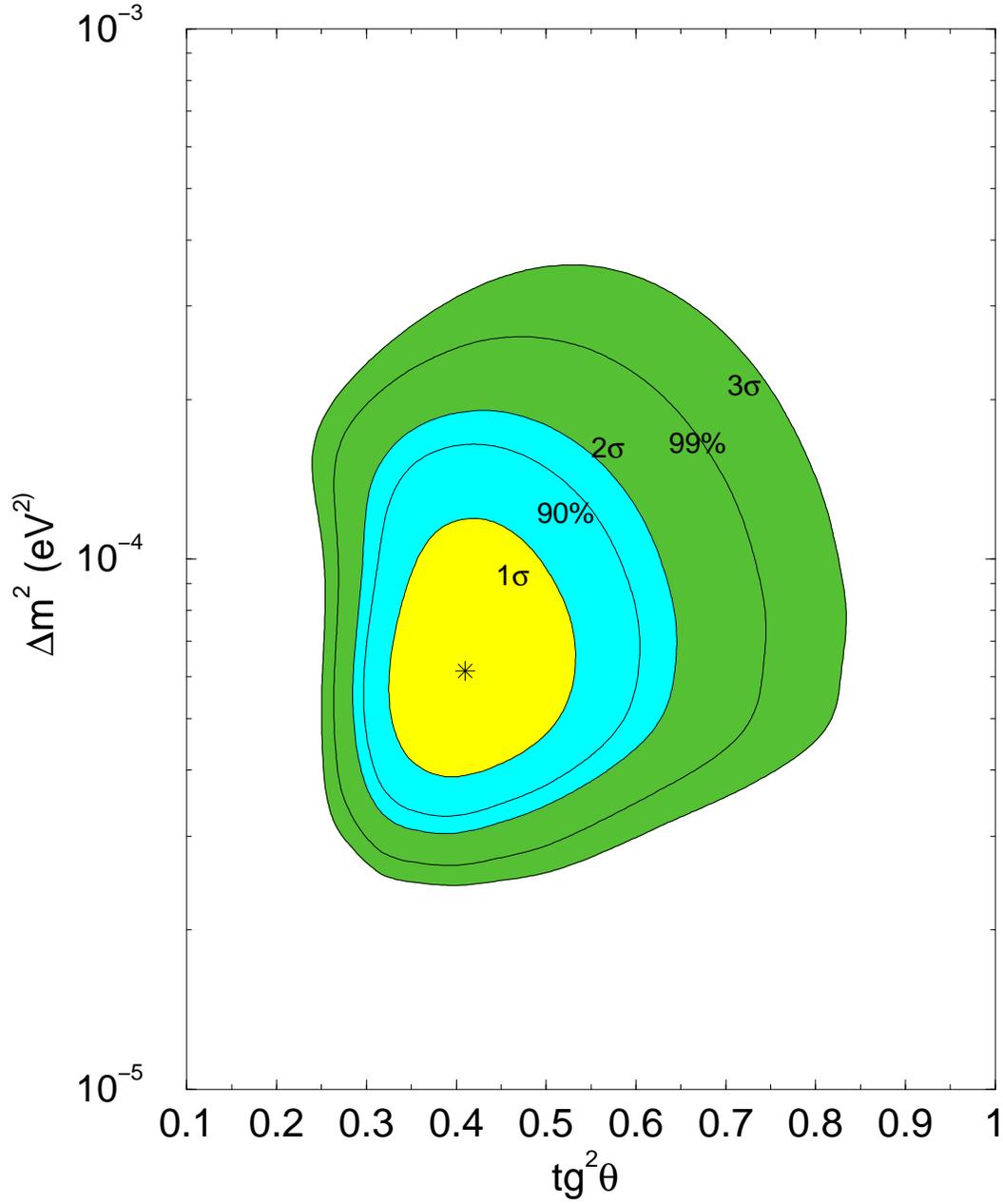}
\caption{
The global  LMA MSW solution. The  boron 
neutrino flux is considered as free parameter. 
The best fit point is marked by a star.  
The allowed regions   are shown at 1$\sigma$, 90\% C.L., 2$\sigma$, 
99\% C.L. and 3$\sigma$.  
}
\label{globallma} 
\end{figure}

The CL contours (see fig.~\ref{globallma})  
shrink substantially as compared with previous determination 
\cite{barger,fogli,valencia,calcutta,plamen,gago,torr}.

From fig.~\ref{globallma} we find the following bounds on oscillations
parameters: 

1)  $\Delta m^2$ is rather sharply restricted from below 
by the day-night asymmetry at SK: 
$\Delta m^2 > 2.3 \cdot 10^{-5}$  eV$^2$ at 99.73\%  {\rm C.L.} . 

2) The upper limits  on $\Delta m^2$ for different confidence levels 
equal:  
\be
\Delta m^2 \leq \left\{
\begin{array}{ll}
1.2 \times 10^{-4} ~ {\rm eV}^2,   &~~ 68.27\% ~ {\rm C.L.}\\ 
1.9 \times 10^{-4} ~ {\rm eV}^2,   & ~~95\% ~ {\rm C.L.}\\
3.6 \times 10^{-4} ~ {\rm eV}^2,   & ~~99.73\% ~ {\rm C.L.} \\ 
\end{array}
\right. . 
\label{up-on-dms}
\ee
All these limits are stronger than the 
CHOOZ~\cite{CHOOZ} bound which appears for maximal 
mixing at $\Delta  m^2 \sim 8\times10^{-4} {\rm eV}^2$.  

3) The upper limit on mixing angle becomes substantially stronger than
   before: 
\be
\tan^2 \theta < \left\{
\begin{array}{ll}

  0.53~~  &  68.27 \% ~~    {\rm C.L.} \\ 
  0.65~~  &  95 \% ~~    {\rm C.L.} \\ 
  0.84~~  &  99.73 \%  ~~   {\rm  C.L.}\\
\end{array}
\right.
\label{theta-up}
\ee 
Maximal mixing is allowed  
at $\sim 4 \sigma$ level for $\Delta m^2 = (5 - 7) \cdot 10^{-5}$
eV$^2$.  

\begin{figure}[ht]
\centering\leavevmode
\epsfxsize=.8\hsize
\epsfbox{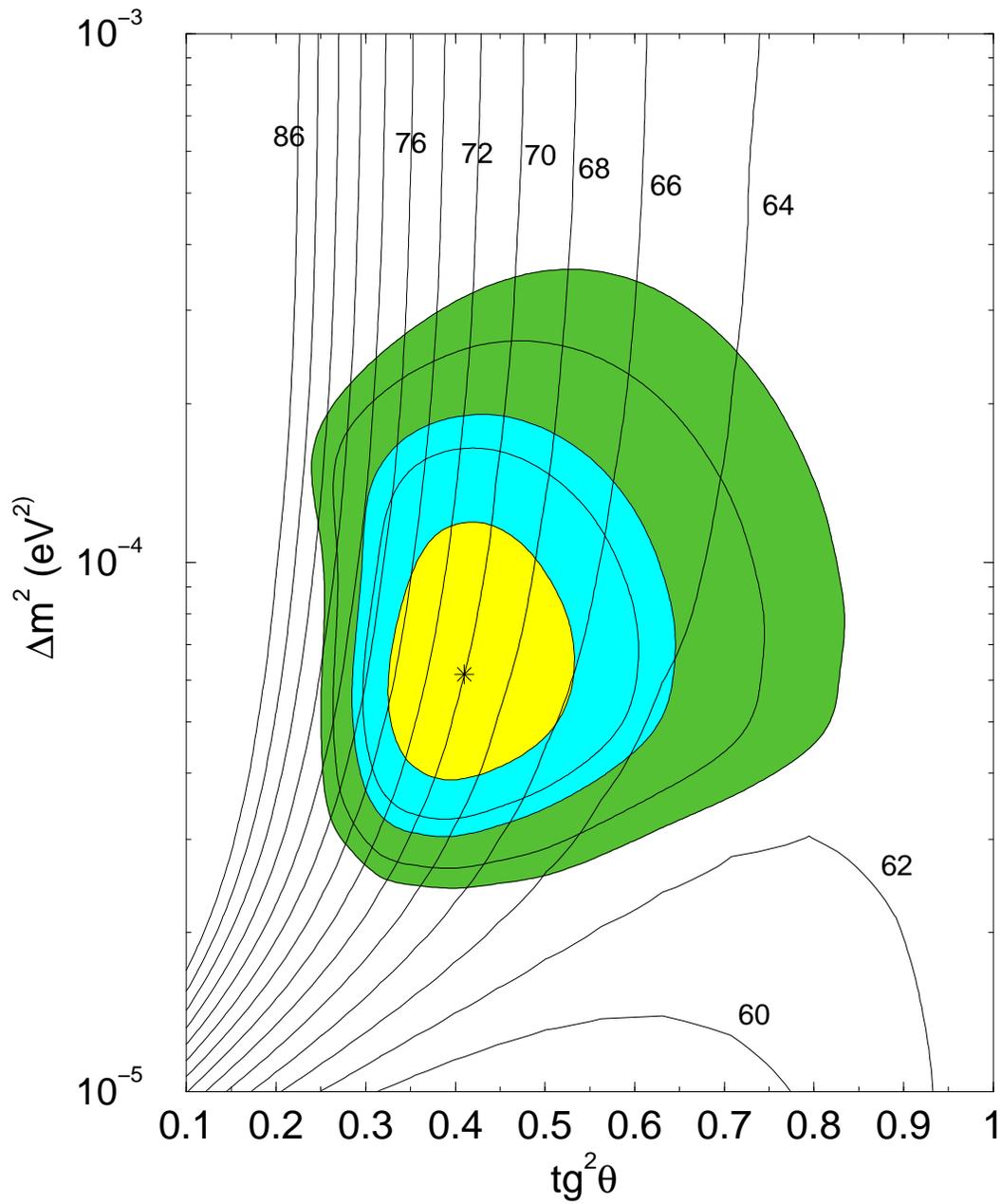}
\caption{
Lines of constant $Ge$-production rate (number at the curves in SNU) 
in the LMA region. In the best fit point: $R_{Ge}=70.5$ SNU.
}
\label{galliumlma} 
\end{figure}

\begin{figure}[ht]
\centering\leavevmode
\epsfxsize=.8\hsize
\epsfbox{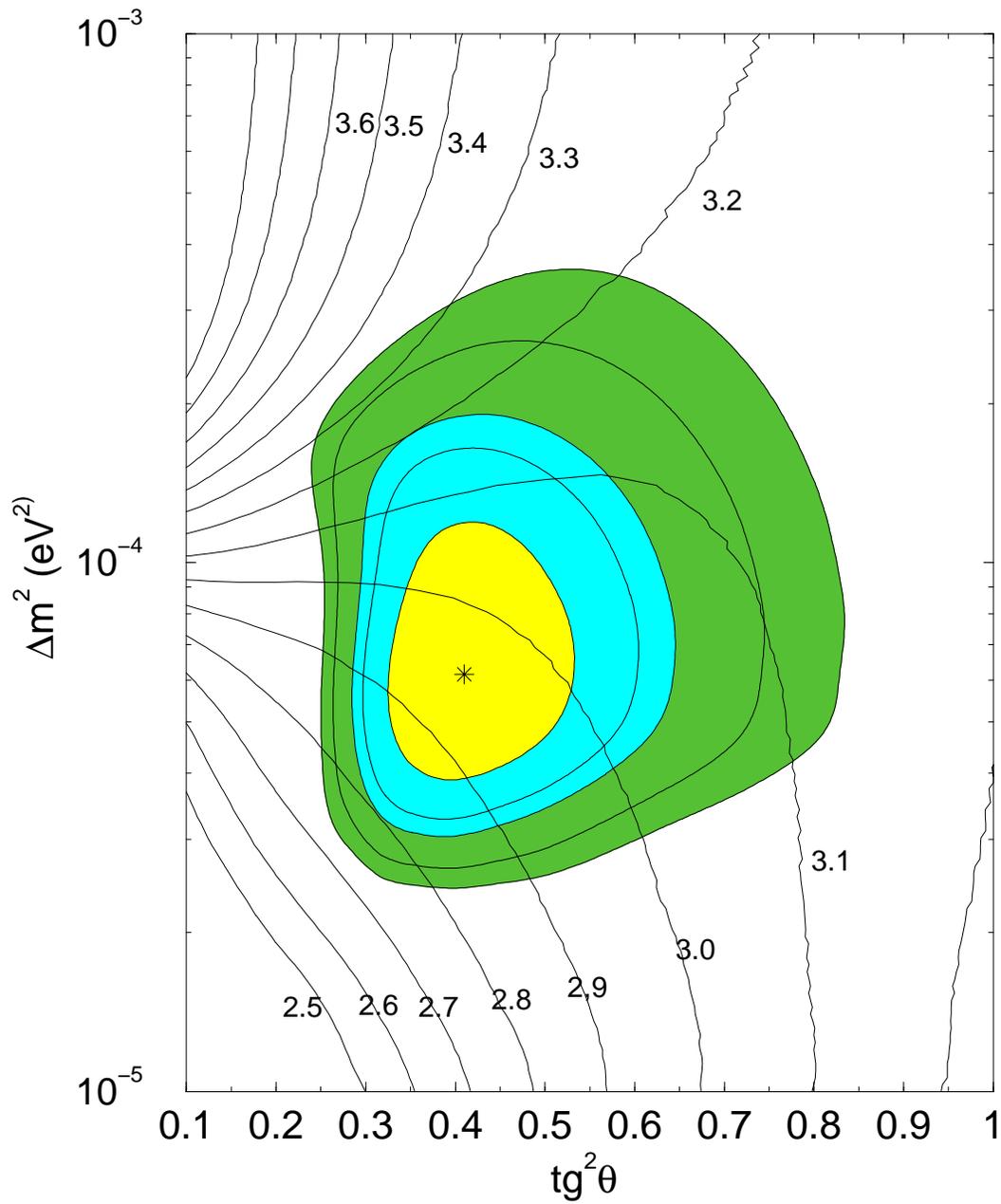}
\caption{
The same as in Fig~\ref{galliumlma}, but for the Ar-production rate.
In the best fit point: $R_{Ar}=2.95$ SNU. The dependence of $f_B$ on
oscillation parameters is taken into account.
}
\label{chlorinelma} 
\end{figure}

Notice that the SNO data alone exclude  maximal mixing at 
about 3$\sigma$: the data determine now rather precisely 
the NC/CC ratio which is directly related to $\sin^2 \theta$. 
Also observed Germanium production rate as well as Argon production  
rate disfavor maximal mixing (see fig.~\ref{galliumlma} and 
\ref{chlorinelma}).

So, now we have strong evidence that solar neutrino mixing significantly 
deviates from maximal value. One can introduce the deviation
parameter~\cite{max} 
\be
\epsilon \equiv 1 - 2\sin^2\theta . 
\ee 
From our analysis we get 
\be
\epsilon  > 0.08,~~~(3 \sigma).
\ee 
That is, at $3 \sigma$: 
$\epsilon > \sin^2\theta_c$,  where $\theta_c$ is the Cabibbo angle. 
This result has important theoretical implications.

4) {\it lower} limit on  mixing : 
\be
\tan^2 \theta > 0.23,~~~~   99\% ~{\rm  C.L.}. 
\label{theta}
\ee 
is changed weakly.

\begin{figure}[ht]
\centering\leavevmode
\epsfxsize=.8\hsize
\epsfbox{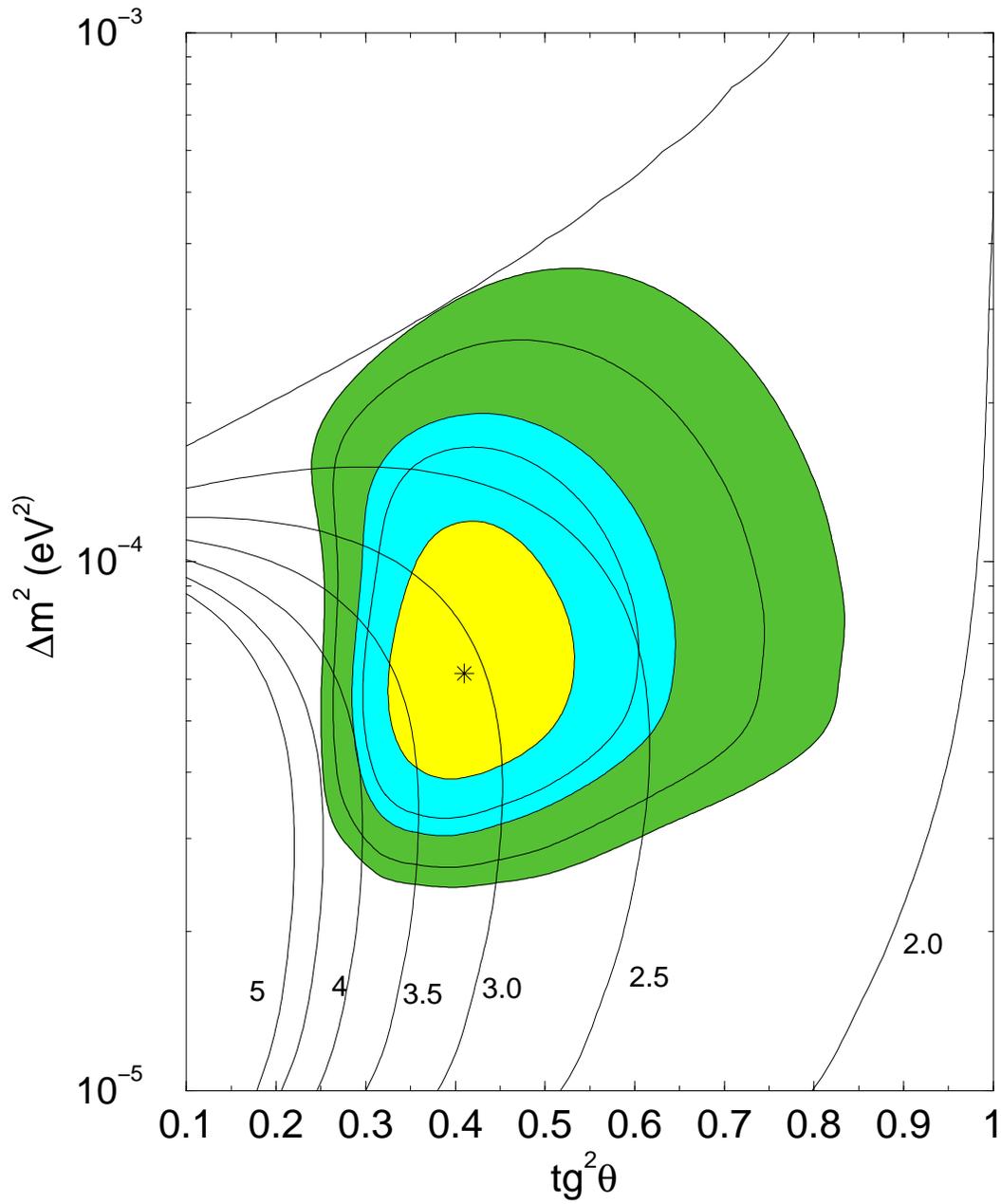}
\caption{
Lines of constant NC/CC ratio in the LMA region. 
In the best fit point: $NC/CC=3.15$.
}
\label{nccc_lma} 
\end{figure}

\begin{figure}[ht]
\centering\leavevmode
\epsfxsize=.8\hsize
\epsfbox{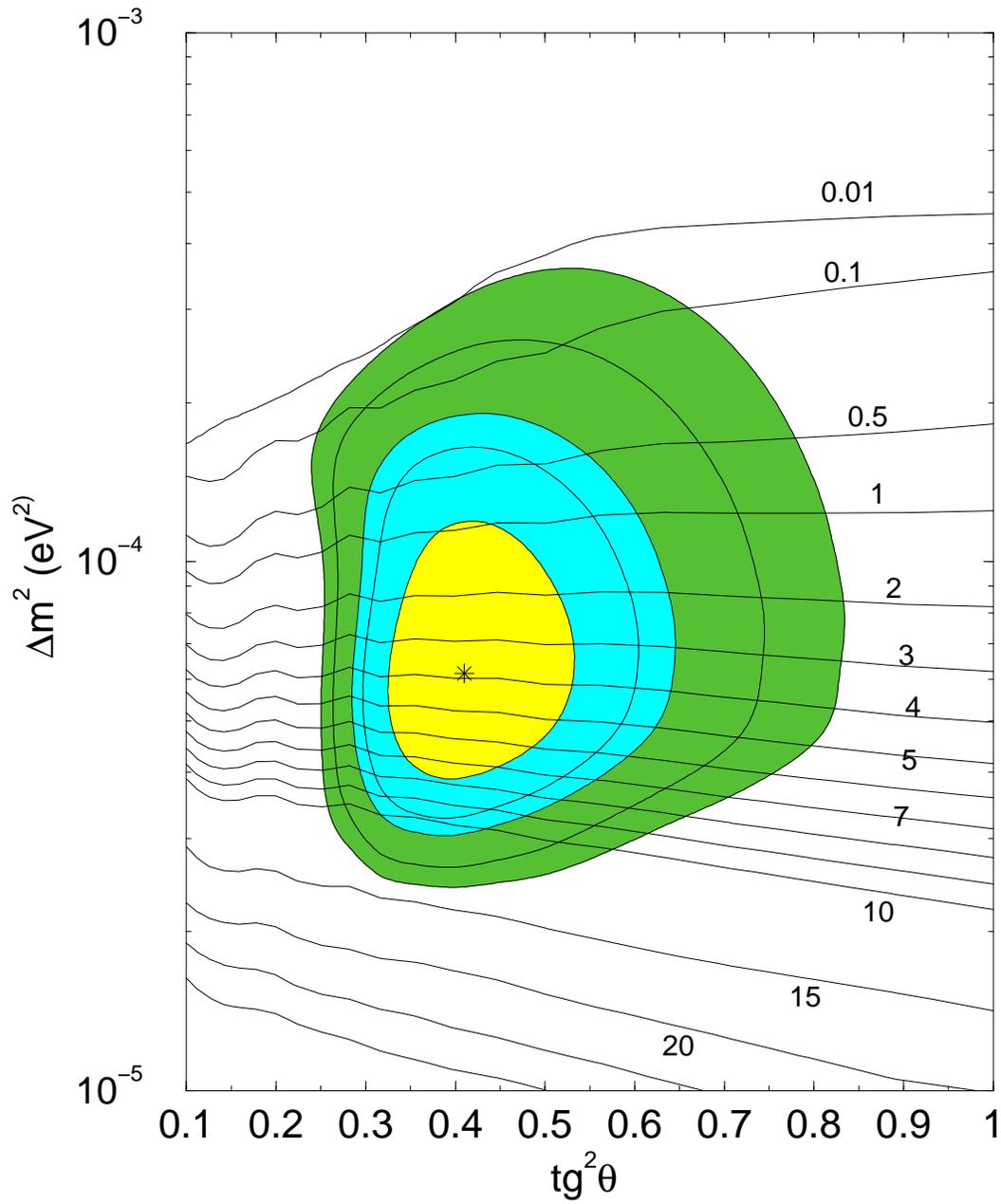}
\caption{
Lines of constant day-night asymmetry of CC events.  
In the best fit point: $A_{DN}^{CC}=3.9\%$.
}
\label{adn_lma} 
\end{figure}

In fig.~\ref{galliumlma}-\ref{adn_lma} we show the grids of predicted
values for various observables.

According to  the pull-off diagram and
figs.~\ref{galliumlma}-\ref{adn_lma}, the LMA solution reproduces 
observables at $\sim 1\sigma$ or better.
The largest deviation is for
the $Ar-$production rate: the solution predicts $1.6\sigma$ 
larger rate than the Homestake result. 

The best fit point value and $3\sigma$ interval 
for Ge production rate  equal 
\be
Q_{Ge} = 70.5~SNU, ~~~ Q_{Ge} = (63 - 84) SNU, ~~~~~  3\sigma .  
\ee
Notice that at maximal mixing $Q_{Ge} < 63$ SNU which is 
$2\sigma$ away from the combined experimental result.

\subsection{VAC}

In the best fit point we get $\chi^2 = 74.9$ and: 
\be
\chi^2(VAC) -  \chi^2(LMA) = 9.7 . 
\ee
So, this solution is accepted at $3\sigma$ level. 
Notice that the solution appears in the dark side of the parameter space 
which means that some matter effect is present. 
This solution was ``discovered'' in 1998 and its properties 
have already been described in the literature.
Clearly it does not predict any day-night asymmetry. 
The solution requires rather low ($1.6\sigma$)  Boron neutrino flux
and gives rather poor description of rates (see fig.~\ref{pulloff}). In
particular, 2.7$\sigma$ higher Ar-production rate and 2.6$\sigma$
lower NC/CC ratio are predicted. 
Imposing the SSM restriction on this flux 
leads to exclusion of this VAC solution at $3\sigma$ level.

\subsection{Any chance for SMA?}

We find that the best fit point from the  SMA region 
has $\chi^2 = 99.8$. For the difference of $\chi^2$ we have:      
\be 
\chi^2(SMA) - \chi^2(LMA) = 34.5 .
\ee
That is, SMA is accepted at  $5.5\sigma$ only. Moreover, the 
solution requires about $3\sigma$ lower boron neutrino 
flux than in the SSM.   
It  predicts negative Day-Night asymmetry: 
$A_{DN}^{CC} = - 0.93\%$. 

Our results are in qualitative agreement with those
in~\cite{barger-a}, where even larger $\Delta \chi^2$ has been
obtained. 

We find that the $\chi^2$ increases weakly with 
$\tan^2 \theta$ up to $\tan^2 \theta = 1.5 \cdot 10^{-3}$,  where 
$\chi^2 \sim 105$. 

Is SMA excluded? We find that very bad fit is due to 
latest SNO measurements of day and night spectra. 
We have checked that the analysis of    
the same set of data but CC 
rate from SNO only   (2001 year) instead of spectrum leads to the best fit 
values $\Delta m^2 =  4.8  \cdot 10^{-6}~ {\rm eV}^2$,  
$\tan^2 \theta = 3.9 \cdot 10^{-4}$ and $f_B = $ 
and $\chi^2(SMA) - \chi^2(LMA) = 11$ in a  good agreement with 
results of similar analysis in \cite{smy2}.  
Since CC SNO data are in a good agreement with new NC/CC result, 
just using the NC/CC does not produce substantial change of quality  
of the SMA fit \cite{bahcall-a}.  
So it is the spectral data which give large contribution to 
$\chi^2$.

The SMA solution with very small mixing provides rather
good description of the SK data: the rate and spectra.
The (reduced) rate $R \equiv OBS/SSM$  of the ES events can be
written as
\be
[ES] = f_B [ P_{ee} (1-r) + r],
\ee
where $r \approx 0.16$ is the ratio of $\nu_{\mu} -e$ to $\nu_e - e$
cross-sections. Taking $R_{ES} = 0.45$ and $f_B = 0.58$ we find
the effective  survival probability: $P_{ee} = 0.73$.
Then, for reduced CC event rate we get $[CC] = f_BP_{ee}= 0.425$ 
- close to the ES rate, and moreover,
\be
NC/CC \approx 1/P_{ee} = 1.37
\ee
which is substantially smaller then the observed quantity (\ref{nc/cc}).
So, one predicts in this case a suppressed contribution of the NC events to
the total rates. Correspondingly, significant distortion of the
energy spectrum of events is expected 
with  smaller than observed rate at low energies and
higher rate at
high energies.

This problem with SNO could be avoided for larger
mixing: $\tan^2 \theta > 1.5 \cdot 10^{-3}$ 
(in fact, imposing the SSM restriction on the boron neutrino flux leads
to the shift of the  best fit point to larger $\theta$).
In this case, however serious
problems with SK data appear,  namely, with spectrum distortion and
zenith angle distribution. Strong day-night asymmetry is predicted for
the Earth  core-crossing bin. Previous analysis which used SK day and
night spectra could not realize the latter problem.

Notice that the SNO data alone do not disfavor SMA with 
large $\tan^2 \theta = (1.5 - 2) \cdot 10^{-3}$. This region, 
however is strongly disfavored by SK.

Zenith angle distribution can give a decisive check 
of the SMA solution. The SNO night data could be divided into 
two bins: ``mantle" and ``core". Concentration of the  night excess of
rate in the core bin  \cite{core-bin}  due to parametric enhancement of
oscillations for the core crossing trajectories \cite{par-res},   
would be the evidence of the SMA solution with relatively 
large mixing:  $\tan^2 \theta = (1.5 - 2) \cdot 10^{-3}$.
However, the SK zenith spectra do not show any excess of the 
``core" bin  rate which testify against this possibility. 

Probably  some unknown  systematics could improve the SMA fit. 
Otherwise, this solution is practically excluded.

\subsection{LOW starts to  disappear?}

In  the best fit point   we get $\chi^2 = 78.9$,  so that 
\be
\chi^2(LOW) -  \chi^2(LMA) = 12.4
\ee
which is slightly  beyond the $3\sigma$ range. In contrast with 
other analyses, LOW does not appear at $3\sigma$ level. 
Notice that in the SNO analysis \cite{sno-dn}  the LOW solution exists 
marginally at $3\sigma$ level. Inclusion of 
the SK data which contain information about zenith angle distribution 
(zenith spectra) worsen the fit (this effect has also been observed in
\cite{valencia}).

The LOW solution gives rather poor fit of total rates. 
In the best fit point we get $2.1 \sigma$ larger 
$Ar-$production rate and  $1.2 \sigma$ lower
$Ge-$production rate. 
For the day-night asymmetry of the CC-events we predict 
$A_{DN}^{CC} = 3.5 \%$  and for ES  events: $A_{DN}^{CC} = 2.7 \%$.  
\\

\section{Three neutrino mixing: effect of $\theta_{13}$}

\begin{figure}[ht]
\centering\leavevmode
\epsfxsize=.8\hsize
\epsfbox{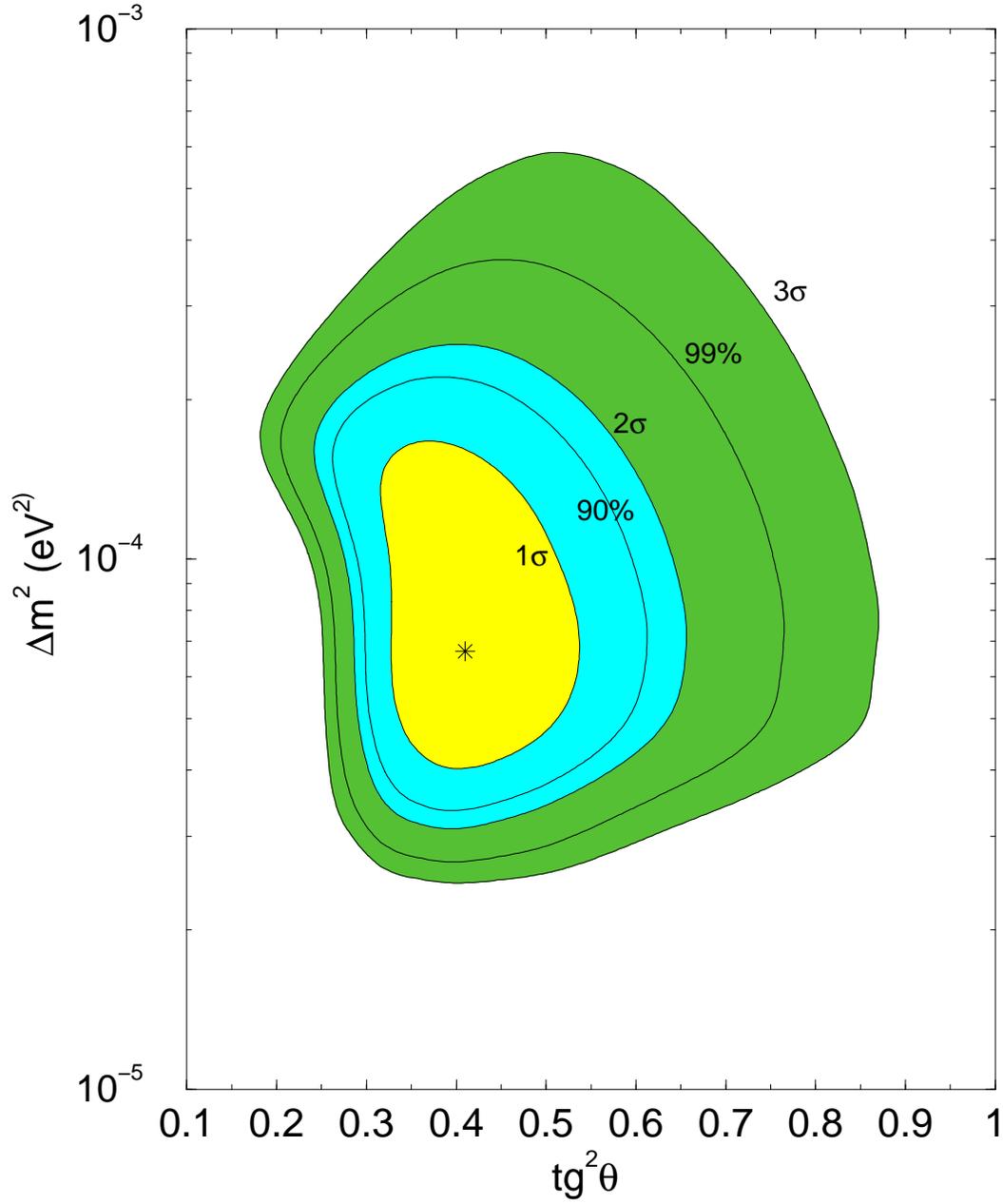}
\caption{
Global LMA solution for 
$\sin^2\theta_{13}=0.04$. 
The  boron 
neutrino flux is considered as free parameter. 
The best fit point is marked by a star.  
The allowed regions   are shown at 1$\sigma$, 90\% C.L., 2$\sigma$, 
99\% C.L. and 3$\sigma$.  
}
\label{globallmaue3} 
\end{figure}

Results of the global analysis in the three neutrino context are shown in
fig.~\ref{globallmaue3}.
To illustrate the effect of third neutrino
we use the three neutrino survival probability (\ref{surv3}) for fixed
value $\sin^2 \theta_{13} = 0.04$ near the upper bound from the 
CHOOZ experiment \cite{CHOOZ}. The number of degrees of freedom is the
same
as in the previous analysis.

We find the best fit point:
\be
\Delta m^2_{12} = 6.7 \cdot 10^{-5} {\rm eV}^2 , ~~~~ \tan^2\theta  = 0.41, ~~~~
f_B = 1.09
\label{global3nu}
\ee
with $\chi^2 = 66.2$. The best fit value of $\Delta m^2_{12}$ is
slightly
higher than that 
in the two neutrino case, whereas the mixing angle is unchanged.
The solution requires slightly higher value of the boron neutrino flux.
 The changes are rather small, however, as a tendency, we find  that
with increase of
$\theta_{13}$ the fit becomes worser in comparison with $2\nu-$ case.
For $\sin^2 \theta_{13} = 0.04$ we get  $\Delta \chi^2 = 1.0$.

In fig.~\ref{globallmaue3} we show the contours of constant confidence 
level constructed
with respect to the best fit point (\ref{global3nu}).
The contours changed  weakly for low mass values
$\Delta m^2_{12} < 10 ^{-4}$  ${\rm eV}^2$ and there are significant changes
for
$\Delta m^2_{12} > 10 ^{-4}$  ${\rm eV}^2$. In particular, the $3\sigma$
upper bound on  $\Delta m^2_{12}$ is $\Delta m^2_{12} < 5.8 \cdot 10
^{-4}$  ${\rm eV}^2$; the 
lower $3\sigma$ bound on mixing:  $\tan^2\theta_{12}  = 0.18$ (compare with
numbers in the
Table 1). Notice, however, that changes are substantially weaker if the
contours are
constructed with respect to the absolute minimum for 
$\theta_{13} = 0$~(\ref{chi-def}).

The changes can be easily understood from the following analytical
consideration.
 
The contribution of the last  term in the probability~(\ref{surv3}) 
is negligible: 
for largest possible value of $\theta_{13}$ it is below 0.5\%. 
So, we can safely use approximation: 
\be
P_{ee} \approx \cos^4 \theta_{13} P_{ee}^{(2)} 
\approx (1 - 2 \sin^2 \theta_{13}) P_{ee}^{(2)}.
\ee  
Mainly the effect of $\theta_{13}$  is reduced to overall
suppression of the survival probability. The  suppression  
factor can be as small as 0.90 - 0.92. 

In the fit with the free boron neutrino flux, the observables 
at high energies ( $>$ 5 MeV) are determined by the  following 
reduced rates
\[
[NC] \equiv \frac{NC}{NC^{SSM}}  =~~  f_B, 
\]
\[
[CC] \equiv \frac{CC}{CC^{SSM}}  =~~ f_B \cos^4 \theta_{13} P^{(2)}_{ee},
\]
\begin{equation}
[ES] \equiv \frac{ES}{ES^{SSM}}  =~~ f_B [\cos^4 \theta_{13} P^{(2)}_{ee}(1
- r) + r].
\label{reducedrates}
\end{equation}
As far as the fit of experimental data on CC-events are concerned 
(SNO, SK, and partly, Homestake), the effects of $\theta_{13}$ is simply
reduced to renormalization  of the boron neutrino flux: 
\be
f_B \rightarrow \frac{f_B}{\cos^4 \theta_{13} }
\ee
without change of the oscillation parameters $\Delta m_{12}$ and
$\theta_{12}$. 
The dependence of the parameters on $\theta_{13}$  appears 
via the ratios of rates, which do not depend 
on $f_B$. From (\ref{reducedrates}) we find 
\be
\cos^4 \theta_{13} \frac{[NC]}{[CC]} \approx \frac{1}{P^{(2)}_{ee}} 
\ee
\be
\frac{\cos^4 \theta_{13}}{r} 
\left[\frac{[ES]}{[CC]} - (1 - r)\right] \approx \frac{1}{P^{(2)}_{ee}} .
\ee
So, the effect of $\theta_{13}$ is equivalent to decrease
of the ratios [NC]/[CC] and [ES]/[CC]. According to
fig.~\ref{globallmaue3}, this 
shifts the allowed regions to larger $\Delta m_{12}$ and $\theta_{12}$.

For low energy measurements (Gallium experiments), 
sensitive to the pp-neutrino flux, which is known rather well,  
the increase of $\theta_{13}$ should be compensated by  increase of the
survival probability. This may occur due to increase of 
$\Delta m_{12}$ or/and decrease of $\tan^2 \theta_{12}$. 

For $\Delta m^2_{12} < 10 ^{-4}$  ${\rm eV}^2$ the boron neutrino spectrum is 
in the bottom of suppression pit and the low energy neutrinos are on the 
adiabatic edge. In the fit, the increase of $\theta_{13}$ is compensated 
by the increase of $f_B$ and $\Delta m^2_{12}$. 
For $\Delta m^2_{12} > 10 ^{-4}$  ${\rm eV}^2$, the spectrum is in the 
region where conversion is determined mainly by averaged 
vacuum oscillations with some matter corrections: 
$P^{(2)}_{ee} \sim (1 -  0.5 \sin^2 2\theta_{12})$. 
The dependence on $\Delta m^2_{12}$ is very weak which explains 
substantial enlargement of the allowed region to large values of 
$\Delta m^2_{12}$. The effect of 
$\theta_{13}$ can be compensated by  decrease of $\theta_{12}$ 
which explains expansion of the region toward  smaller 
$\tan^2\theta_{12}$.

For LOW solution increase of $\theta_{13}$ leads to improvement of the
fit, so that this solution appears (for $\sin^2\theta_{13}=0.4$) at
3$\sigma$ level with respect to best fit point (\ref{global3nu}).
Also for the SMA solution the effect of  $\theta_{13}$ leads to 
slight improvement  of the fit.  

\section{Predictions for KamLAND}

Next step in developments will be probably related to operation of 
the KamLAND experiment \cite{busenitz}. Both total rate of events above 
the effective threshold $T_{eff}$ and the energy spectrum of events will be measured. 

We characterize the effect of the oscillation disappearance  by 
the ratio of the total number of events with visible energy above $T_{eff}$: 
\be
R_{Kam} = \frac{1}{N_0} \int_{T_{eff}}\int_{T'}\int_E dE dT' dT \sum F_i P_{i} \frac{d\sigma}{dT} f(T, T') \,,
\ee
where $F_i$ is the flux from $i$ reactor, $P_{i}$ is the survival probability 
for neutrinos from $i$ reactor, $\sigma$ is the cross-section 
of the detection reaction,  $f(T, T')$ is the energy resolution. 
$N_0$ is the rate without oscillations ($P_i = 1$). 

In our calculations we used the energy spectra of  reactor neutrinos from
\cite{vogel_kl,murayama_kl}. 
The differential cross-section of  the $p+\overline{\nu_e}\rightarrow
n+e^+$ reaction, is taken from ~\cite{beacom}.   The parameters  of the 16 nuclear
reactors, maximal thermal  power, distance from the reactor
to the detector, etc., are given
in~\cite{busenitz}.  We used the Gaussian form for the energy resolution   function
$f(T,T')$ with $\sigma/E = 5\%/\sqrt{E(MeV)}$,  and  
$T_{eff}=2.6$ MeV as the threshold for the  visible energy ~\cite{shirai}.
   
\begin{figure}[ht]
\centering\leavevmode
\epsfxsize=.8\hsize
\epsfbox{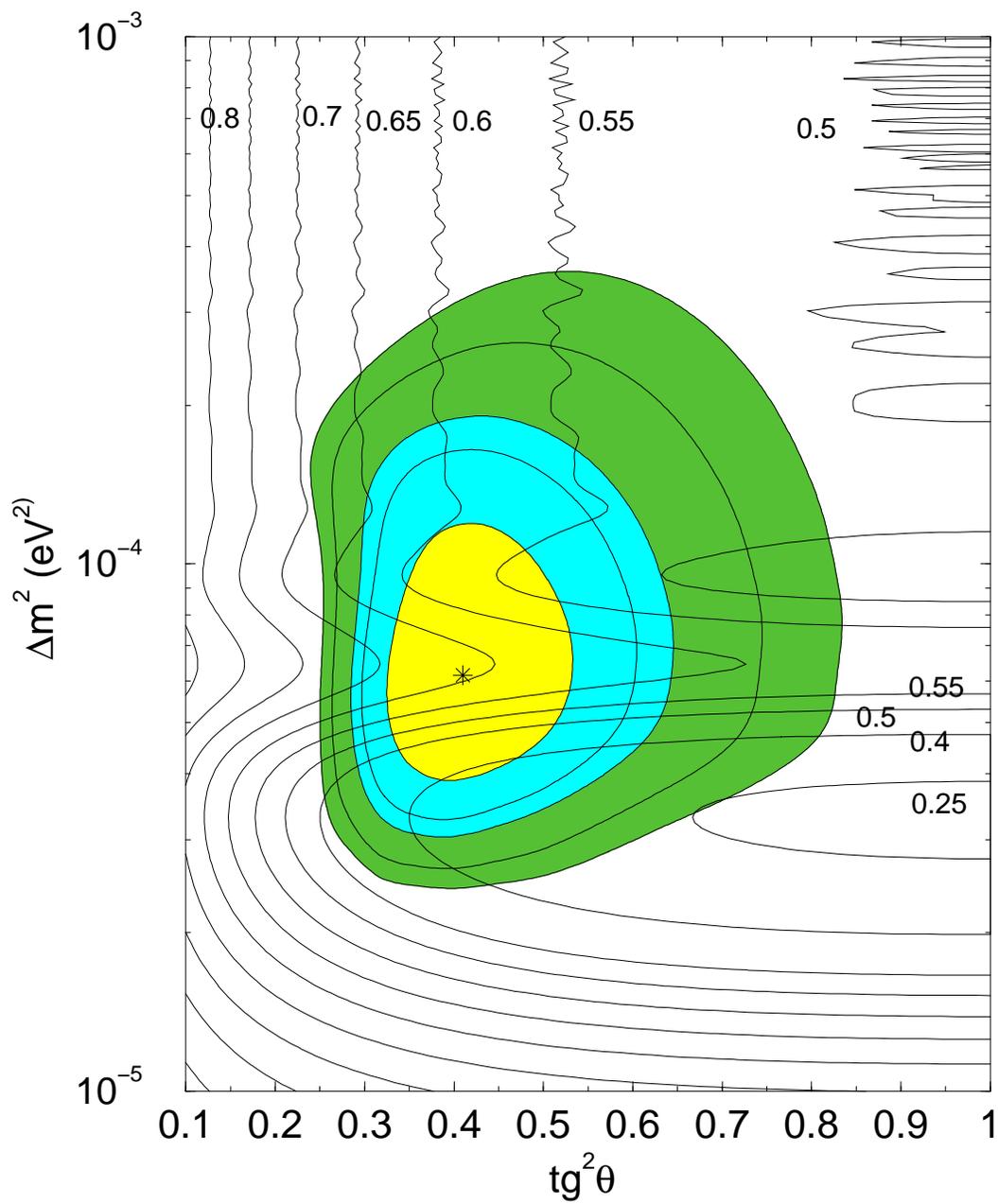}
\caption{
Lines of constant total suppression at KamLAND. In the best fit point: 
$R_{Kam}=0.65$.
}
\label{kamtot} 
\end{figure}

In fig.~\ref{kamtot} we show the contours of  constant suppression factor in the 
$\Delta m^2 - \tan^2 \theta$ plot. In the best fit point 
\be
R_{Kam} = 0.65 \,,
\ee
and in the $1\sigma$ region: $R_{Kam} = 0.4 - 0.7$. 

Notice that the best fit point is in the range of lowest sensitivity of the 
total rate on $\tan^2 \theta$. If {\it e.g.} $R_{Kam}$ is measured 
with $8\%$ accuracy  which would correspond to $R_{Kam} = 0.65 \pm 0.05$, 
we get from the fig.~\ref{kamtot}  that any mixing 
in the interval  $\tan^2 \theta = 0.12 - 1.0$ is allowed. 

The suppression factor strongly depends on   
$\Delta m^2$ in the range below the best fit point and this dependence is
very weak  for $\Delta m^2 > 10^{-4}$ eV$^2$. No bound on $\Delta m^2$ from 
the allowed region can be obtained by measurements of the total rate. 

\begin{figure}[ht]
\centering\leavevmode
\epsfxsize=.8\hsize
\epsfbox{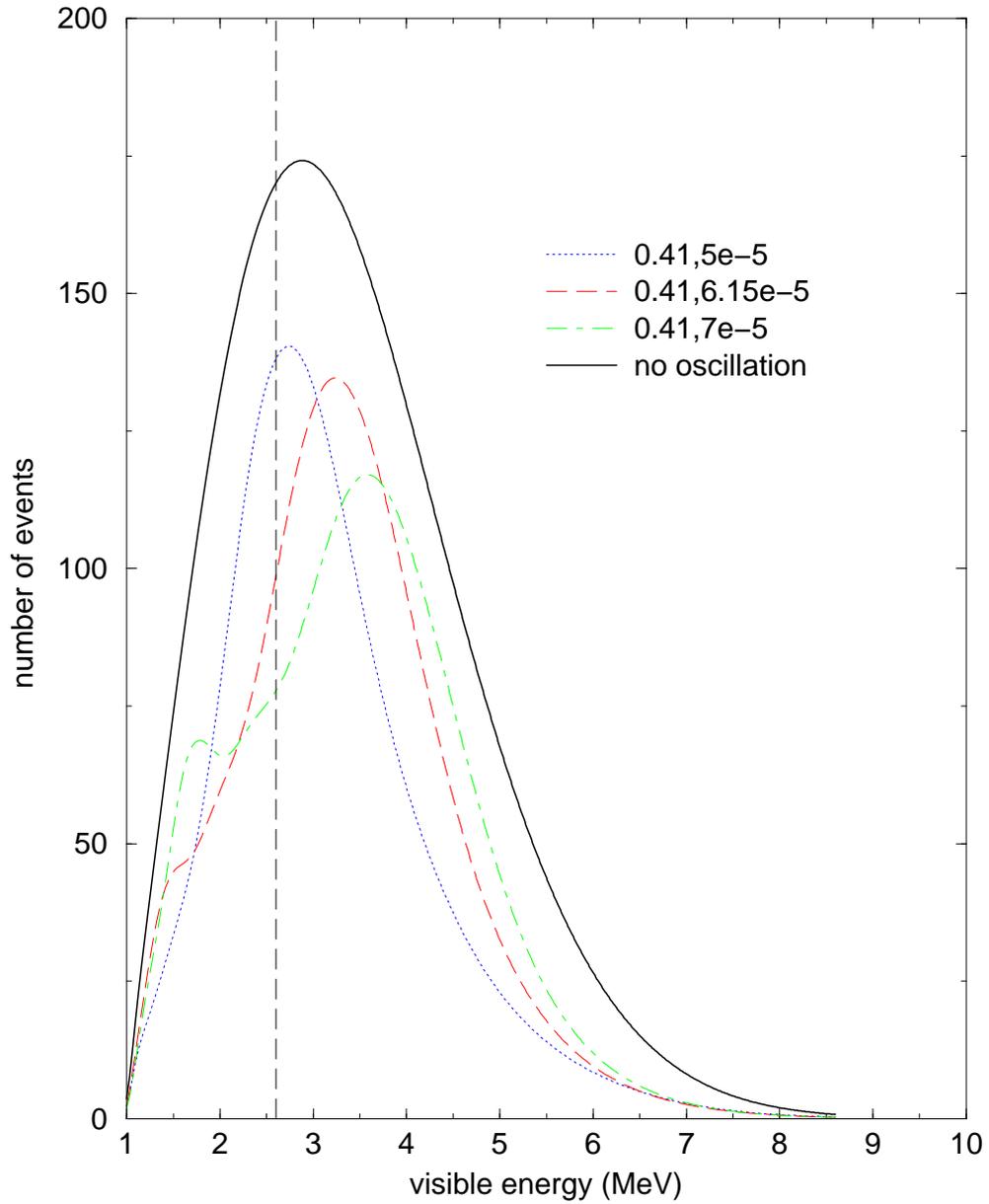}
\caption{
Spectral distortion for three different values on $\Delta m^2$,
including the best fit point of our analysis.
}
\label{kamspe} 
\end{figure}

The distortion of the visible energy spectrum depends on 
$\Delta m^2$ strongly. In fig.~\ref{kamspe} we show the spectrum for
different values of $\Delta m^2$. There is a shift of maximum to
large $E$ with increase of  $\Delta m^2$. For the best fit 
value of $\Delta m^2$ the maximum is at $E \approx 3.5$ MeV. The most
profound 
effect of oscillations is the suppression of rate at high energies. 
For instance, for $E \approx 5$ MeV the suppression 
factor is smaller than 1/2.

\section{Conclusions}

We find that the LMA MSW solution with parameters $\Delta m^2
\sim 6.15 \cdot 10^{-5}$ eV$^2$ and $\tan^2 \theta = 0.41$
gives the best fit to the data.
The solution reproduces well the zenith spectrum measured by SK and
the day and night spectra at  SNO.   
It is in a very good agreement with SSM flux of
the boron neutrino:  $f_B = 1.05$.

The recent SNO results together with zenith spectra results from SK slightly
shifted the best fit point to larger $\Delta m^2$ and $\theta$. 
At the same time the allowed regions of oscillation
parameters shrunk, leading to important, and statistically significant,
upper bounds on mixing angle and $\Delta m^2$. Now we have strong evidence
that ``solar'' mixing is non-maximal, and moreover, deviation from maximal
mixing
is rather large. 
We find that QuasiVacuum oscillation solution with
$\Delta m^2 = 4.5 \cdot 10^{-10}$ eV$^2$ and mixing in the dark side
is the only other solution accepted at $3\sigma$ level,
provided that the boron neutrino flux is about 30\% below the SSM value.

The LOW solution is accepted at slightly higher than $3\sigma$ level
and it reappears at $3\sigma$ level if $\theta_{13}$ is included.

The SMA solution gives very bad fit of the data especially the
SNO spectra predicting rather small contribution of the NC events in
comparison with CC events.

We find that $\theta_{13}$ produces rather small effect on the solutions
even with new high statistics data. As a tendency we see that inclusion of
the $\theta_{13}$ effect worses fit of the data in the LMA region,
and shifts the best fit point to larger $\Delta m^2_{12}$.

We have found predictions for the KamLAND experiment: 
in the best fit point one expects the suppression factor for 
total signal $\sim 0.6 - 0.7$ and the spectrum distortion with 
substantial suppression in the high energy part.


\section*{Acknowledgments}

We thank Prof. Mark Chen for clarification 
of the way SNO treats the correlations between the 
systematic uncertainties. The authors are grateful to J. N. Bahcall 
for emphasizing the necessity to take into account correlations  
of cross-sections uncertainties in contributions of different fluxes
to Ar and Ge-production rates we have discussed in sect. 2.4.

\section*{Appendix}

The systematics uncertainties were treated according 
to~\cite{fogli95}. Writing the total counting rate in SNO as a sum
over different contributions and different spectral bins, we have:

\begin{equation}
R_j = \sum_{i=1,5} R_{ij} \,,
\label{eq:rj}
\end{equation}
where the index $j$ stands for the different spectral bins and $i$
runs over the five contributions to the SNO data (CC, NC, ES, neutron
background and low energy background). We assume that all
systematic uncertainties of the SNO result  are the uncertainties in the
theoretical
prediction. These uncertainties can be written in terms of the
systematics uncertainties of the input parameters of experiment ($X_k$):
\begin{equation}
\label{eq:sigma1}
\sigma^2_{j_1,j_2}(TH)=\sum_{k=1,14}
\frac{\partial R_{j_1}}{\partial ln X_k}
\frac{\partial R_{j_2}}{\partial ln X_k} (\Delta ln X_k)^2 .
\end{equation}

We take the  systematic uncertainties from~\cite{sno-how}. 
The different systematic uncertainties are added
 in quadrature.

Eq.(\ref{eq:sigma1}) can be written in terms of the different
contributions $R_{ij}$ ~(\ref{eq:rj}) as:
\begin{equation}
\sigma^2_{j_1,j_2}(TH)=\sum_{i_1=1,5}\sum_{i_2=1,5}
R_{i_1j_1}R_{i_2j_2}
\sum_{k=1,14} \alpha_{i_1,j_1,k}\alpha_{i_2,j_2,k}(\Delta ln X_k)^2 \,,
\label{eq:sigma2}
\end{equation}
where we have introduced the parameters $\alpha_{i,j,k}$:
\begin{equation}
\alpha_{i,j,k} \equiv \frac{\partial ln R_{i,j}}{\partial ln X_k}.
\end{equation}

These parameters are numerically 
estimated by changing the response function of
the detector through changes in the parameters $X_k$.


\newpage

\newpage

\newpage

\newpage

\newpage

\newpage

\newpage

\newpage

\newpage

\end{document}